# A Galactic-scale gas wave in the Solar Neighborhood

João Alves[1,2], Catherine Zucker[3], Alyssa A. Goodman[2,3], Joshua S. Speagle[3], Stefan Meingast[1], Thomas Robitaille[4], Douglas P. Finkbeiner[3,5], Edward F. Schlafly[6], and Gregory M. Green[7]

[1] University of Vienna, Department of Astrophysics, Türkenschanzstrasse 17, 1180 Wien, Austria
[2] Radcliffe Institute for Advanced Study, Harvard University, 10 Garden Street, Cambridge, MA 02138, USA
[3] Harvard University Department of Astronomy, Center for Astrophysics | Harvard & Smithsonian, 60 Garden St. Cambridge, MA 02138, USA
[4] Aperio Software, Headingley Enterprise & Arts Centre, Bennett Road, Leeds LS6 3HN, UK
[5] Harvard University Department of Physics, 17 Oxford Street, Cambridge, MA 02138, USA
[6] Lawrence Berkeley National Laboratory, One Cyclotron Road, Berkeley, CA 94720, USA
[7] Kavli Institute for Particle Astrophysics and Cosmology, Stanford University, 452 Lomita Mall, Stanford, CA 94305, USA

**For the past 150 years, the prevailing view of the local Interstellar Medium (ISM) was based on a peculiarity known as the Gould's Belt[1, 2, 3, 4], an expanding ring of young stars, gas, and dust, tilted about 20° to the Galactic plane. Still, the physical relation between local gas clouds has remained practically unknown because the distance accuracy to clouds is of the same order or larger than their sizes[5, 6, 7]. With the advent of large photometric surveys[8] and the Gaia satellite astrometric survey[9] this situation has changed[10]. Here we report the 3-D structure of all local cloud complexes. We find a narrow and coherent 2.7 kpc arrangement of dense gas in the Solar neighborhood that contains many of the clouds thought to be associated with the Gould Belt. This finding is inconsistent with the notion that these clouds are part of a ring, disputing the Gould Belt model. The new structure comprises the majority of nearby star-forming regions, has an aspect ratio of about 1:20, and contains about 3 million solar masses of gas. Remarkably, the new structure appears to be undulating and its 3-D distribution is well described by a damped sinusoidal wave on the plane of the Milky Way, with an average period of about 2 kpc and a maximum amplitude of about 160 pc. Our results represent a first step in the revision of the local gas distribution and Galactic structure and offer a new, broader context to studies on the transformation of molecular gas into stars.**

To reveal the physical connections between clouds in the local ISM, we determined the 3-D distribution of all local clouds complexes[11] by deriving accurate distances to 380 lines of sight. The lines of sight were chosen to include not only all known local clouds[10, 12] but also potential bridges between them, as traced by lower column density gas. Figure 1 presents the distribution of lines of sight studied towards the Galactic anti-center and illustrates our overall approach. Each line of sight covers an area in the sky of about 450 arcmin$^2$ and includes both foreground and background stars to a particular cloud. The distances and the colors of these stars are used to compute a distance to the cloud (see the Methods section).

In Figure 2 (interactive) we present the distribution of cloud distances to all of the studied lines of sight in a Cartesian XYZ frame where X increases towards the Galactic center, Y increases along the direction of rotation of the Galaxy, and Z increases upwards out of the Galactic plane. In the XY projection (a top-down view of the Galactic disk), it is clear that cloud complexes are not randomly distributed but instead tend to form elongated and relatively linear arrangements. Surprisingly, we find that one of the nearest structures, at about 300 pc from the Sun at its closest point, is exceptionally straight and narrow in the XY plane. This straight structure (1) undulates systematically in the Z-axis for about 2.7 kpc on the XY plane, (2) is co-planar for essentially its entire extent, and

(3) displays radial velocities[13] indicating that the structure is not a random alignment of molecular cloud complexes but a kinematically coherent structure. We find that this new structure is well-modeled as a damped sinusoidal wave. The red points in Figure 2 were selected by the fitting procedure, by explicitly modeling inliers and outliers. We tested the validity of the model by modeling the tenuous connections separately and confirming they meet the same inlier criteria first applied to the major clouds. For more details on the statistical modeling, see the Extended Data section.

Apart from the continuous undulating 3-D distribution, there is also very limited kinematic evidence that the structure is physically oscillating around the mid-plane of the Galaxy, as any sinusoidal mass distribution centered on the Galactic plane should. The Galactic-space U-V-W velocities in the Local Standard of Rest (LSR) frame for a sample of young stellar objects associated with the Orion A cloud near the "trough" of this structure are (-10.2, -1.2, -0.1) km/s (J. Großschedl, private communication), implying that Orion A has now reached maximum distance from the Galactic plane before falling back into it. These observations also indicate that Orion, and likely the large structure described here, is moving tangentially with about the same speed as the local Galactic disk.

This spatially and kinematically coherent structure, that from hereon we call the Radcliffe Wave, has an amplitude of roughly 160 pc at its maximum and a period of roughly 2 kpc. We estimate the mass of the structure to be at least $3 \times 10^6$ $M_\odot$ by integrating the Planck opacity map[14] for the different cloud complexes in the new structure at their estimated distances. The procedures used to compute the mass and model the 3-D shape of the structure are described in the Methods section. We name the new structure the Radcliffe Wave in honor of both the early 20th century female astronomers from Radcliffe College and the interdisciplinary spirit of the current Radcliffe Institute that contributed to this discovery. The new structure can also be seen at lower resolution in recent all-sky 3D dust maps[15, 16, 17] (see Figure 2). A second linear structure, the "split"[16] is about 1 kpc long and seems to contain the Sco-Cen, Aquila, and Serpens clouds, plus a previously unidentified complex. The functional form of the "split" is different, however, in that it is largely confined to the plane over much of its length and does not seem to be undulating.

In Figure 2 one can interactively display the 3-D location of Gould's Belt[18], illustrating that, with the improved distances, this structure is a poor fit to the data, comprising only clouds from Sco-Cen and Orion – the traditional anchors of the Gould's Belt. This fact alone challenges the existence of a Belt, as two points can always define a ring. Because 4/5 of the Gould's Belt clouds (Orion, Perseus, Taurus, Cepheus) are part of the much larger Radcliffe Wave, while 1/5 (Ophiuchus) is part of the "split", we propose that the Gould's Belt is a projection effect of two linear cloud complexes against the sky. Our results provide an alternate explanation for the 20$^\circ$ inclination of the Gould's Belt: it is simply the orientation of the Wave from trough (Orion) to crest (Cepheus). With the benefit of hindsight, the XY distribution of local B-stars in these regions from the 30-year old Hipparcos satellite[19] resembles the two elongated linear structures in Figure 2 more closely than a ring, bolstering previous suspicion that Gould's Belt is a projection effect[20].

In Figure 2, one can interactively display the 3-D location of the Local Arm as traced by masers[21] and investigate the relation between the Radcliffe Wave and the Local Arm. The new structure (red points) is about 20% of the width and 40% of the length of the Local Arm[22], and makes up for an important fraction of the Arm's mass and number of cloud complexes. On the other hand, the Local Arm is much more dispersed and includes local complexes that are not part of the Wave (e.g. Mon OB1, California, Cepheus Far, Ophiuchus, etc.). While there is an excellent agreement between our distance measurements and the maser distances[12], the log-spiral maser fit crosses the new structure at about a 25° angle. The mismatch between the new structure and the log-spiral fit suggests the Local Arm is more structured and complex than previously thought, but is consistent with arms being composed of quasi-linear structures on kpc scales.[23, 24]

The origin of the Radcliffe Wave is unclear. The structure is too large (and too straight) to have been formed by the feedback of a previous generation of massive stars. More likely, this narrow structure is the outcome of a large-scale Galactic process of gas accumulation, either from a shock front in a spiral arm[25] or from gravitational settling and cooling on the plane of the Milky Way[26]. Linear kpc-sized structures, similar to the one presented here, have been seen in nearby galaxies[27] and numerical simulations[24] of spiral-arm formation.

The undulation of the Wave is even harder to explain. The accretion of a tidally stretched gas cloud settling into the Galactic disk could in principle mimic the shape and the damped undulation, but it requires synchronization with Galactic rotation (Orion's $V_{LSR}$ ~ 0 km/s), which is plausible but seems unlikely. Analog kpc-size waves (or corrugations) have been seen in nearby galaxies[28] with amplitudes similar to undulation seen in Figure 2[29], but their origins often call for perturbers. Identifying possible disruption events, their corresponding progenitors, and their relationship to the new structure is a substantial challenge that should be explored.

Our findings call for a revision of the architecture of gas in the Local Neighborhood and a re-interpretation of phenomena generally associated with Gould's Belt, such as the Lindblad ring and the Cas-Tau / α-Per populations, among many others[30]. The Radcliffe Wave provides a new framework for understanding molecular cloud formation and evolution. Follow-up work, in particular on the kinematics of this structure, will provide new insights into the relative roles of gravity, feedback, and magnetic fields in star formation research.

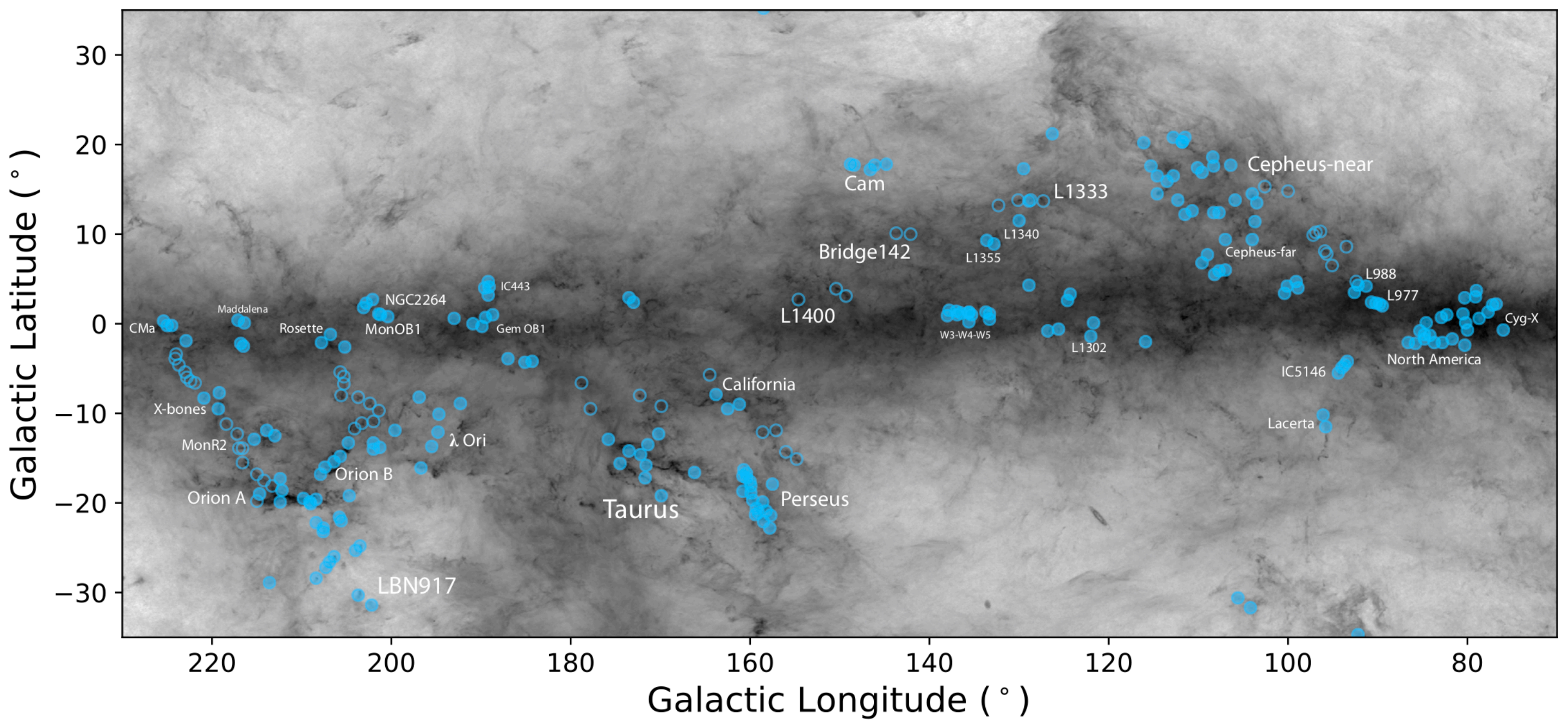

Figure 1: Sky map of targeted star forming regions towards the anti-center of the Milky Way. The filled circles represent the studied lines of sight used to determine accurate distances to known nearby star-forming complexes (the regions labels are roughly proportional to distance). The open circles represent lines of sight toward lower column density envelopes between complexes. The background grayscale map shows the column density distribution derived from Planck data[14].

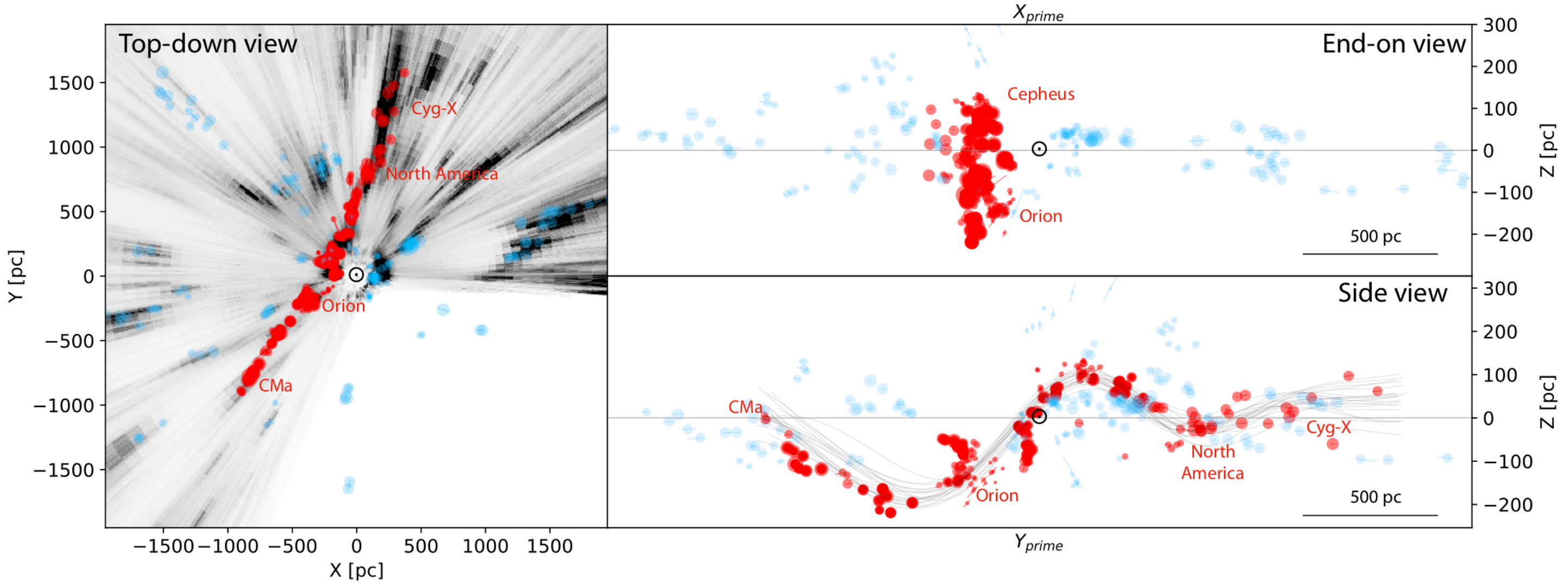

Figure 2: 3-D distribution of local clouds. The position of the Sun is marked with a $\odot$. The size of the symbols is proportional to column density. The red points were selected by a fitting algorithm, as described in the Supplementary methods section. These describe a spatially and kinematically coherent structure we term the Radcliffe Wave (possible models are shown in black in the bottom-right). The grayscale on the XY panel show an integrated dust map[17] ($-300 < Z < 300$ pc), that indicates that our sample of clouds distances is essentially complete. To highlight the undulation and co-planarity of the structure, the right panels show projections in which the XY frame has been rotated anticlockwise by $33^{\circ}$ (top) and clockwise $120^{\circ}$ (bottom) for an observer facing the Galactic anti-center. The $1\sigma$ statistical uncertainties (usually 1 - 2% in distance) are represented by line segments, which are usually smaller than the symbols. There is an additional systematic uncertainty which is estimated to be $\approx 5\%$ in distance[10]. For an interactive version of this figure, including extra layers not shown here (e.g. model of the Gould's Belt, log-spiral arm fits), see https://faun.rc.fas.harvard.edu/czucker/Paper_Figures/radwave.html.

---

## Methods

### Distances.

Distances were determined for 326 sightlines through major local molecular clouds and 54 "bridging" sightlines in-between molecular clouds coincident with the projected structure of the Radcliffe Wave. The methodology used to obtain the distances and the full catalog of sightlines for the major clouds are presented in complementary work[10, 12]. Sightlines for the major clouds were chosen to coincide with star-forming regions in the Star Formation Handbook[11], which is considered to be the most comprehensive resource on individual low- and high-mass star forming regions out to 2 kpc. Sightlines for the tenuous connections were chosen in 2-D to coincide with structures (e.g. diffuse filamentary "bridges"; see Figure 1) which appeared to span the famous star-forming regions on the plane-of-the-sky without a priori knowledge of their distances. These were later used to validate the 3-D modeling, which did not incorporate these distances.

### Mass.

We estimate the mass of the Wave to be about $3 \times 10^6$ $M_{\odot}$ using the Planck column density map shown in Figure 1. To estimate the total mass, we first defined the extent and depth for each complex in Figure 1 using the information on the line-of-sight distances. We then integrated the column density map using the average distance to each complex. To correct for background contamination, which is critical for complexes closer to the plane, we subtracted an average column density per complex estimated at the same Galactic latitude. Our resulting mass estimate of the Wave is likely to be an approximate lower limit to the true mass of the structure as the regions of the wave crossing the plane from Perseus to Cepheus and from Cepheus to Cygnus are poorly sampled due to Galactic plane confusion.

## Kinematics.

We apply the open-source Gaussian fitting package pyspeckit[31] over local $^{12}$CO spectral observations[13] to obtain the observed velocities of the star-forming regions shown in Figure 3. For each sightline, we compute a spectrum over the same region used to compute the dust-based distances. We then fit a single-component Gaussian to each spectrum and assign the mean value as the velocity. We are not able to derive observed velocities to ≈ 25% of the sample that either (1) fall outside the boundaries of the survey[13], (2) have no appreciable emission above the noise threshold, and/or (3) contain spectra that are not well-modeled by a single-component Gaussian. The spectra that are not-well modeled by a single-component Gaussian represent about 2% of sightlines and occur towards the most massive, structured, and extinguished sightlines in the sample, suggesting that these spectra could contain CO self-absorption features. We have confirmed that these more complex spectra do not show evidence of multiple distance components. Regardless, since the predicted velocities rely only on the estimated cloud distances assuming they follow the “universal” Galactic rotation curve[22], not every sightline in Extended Data Figure 1 has a corresponding observed velocity associated with its predicted velocity.

We compute the background grayscale in Extended Data Figure 1 by collapsing the $^{12}$CO spectral observations over only those regions coincident with the cloud sightlines on the plane-of-the-sky.

## 3-D Modeling.

We model the center of the Radcliffe Wave using a quadratic function in X, Y, and Z specified by three sets of “anchor points” $(x_0,y_0,z_0)$, $(x_1,y_1,z_1)$, and $(x_2,y_2,z_2)$. We find that a simpler linear function is unable to accurately model the observed curvature in the structure and subsequently disfavored by the data.

The undulating behavior with respect to the center is described by a damped sinusoidal function relative to the XY plane with a decaying period and amplitude, which we parameterize as

$$\Delta z(t) = A \times \exp\left[-\delta\left(\frac{d(t)}{\mathrm{kpc}}\right)^2\right] \times \sin\left[\left(\frac{2\pi d(t)}{P}\right)\left(1+\frac{d(t)/d_{\max}}{\gamma}\right)+\phi\right] \quad (1)$$

where $d(t) = ||(x,y,z)(t) - (x_0,y_0,z_0)|| = \sqrt{(x-x_0)^2+(y-y_0)^2+(z-z_0)^2}$ is the Euclidean distance from the start of the wave as parameterized by t, $d_{max}$ is the distance at the end of the wave, A is the amplitude, P is the period, $\phi$ is the phase, $\delta$ sets the rate at which the amplitude decays, and $\gamma$ sets the rate at which the period decays. We explored introducing an additional parameter to account for rotation around the primary axis determined by our quadratic fit, but found the results were entirely consistent with the structure oscillating in the XY plane and so excluded it in our final model.

We assume the distance of each cloud $d_{cloud}$ relative to our model to be Normally distributed with some unknown scatter σ, roughly equivalent to the radius of the Wave. To account for different positions along the wave, we take the distance to be defined relative to the closet point such that

$$d_{\rm cloud} = \min_t \left( ||(x_{\rm cloud}, y_{\rm cloud}, z_{\rm cloud}) - (x_{\rm wave}, y_{\rm wave}, z_{\rm wave})(t)|| \right) \quad (2)$$

Finally, we account for structure "off" the Wave by fitting a mixture model. We assume that some fraction *f* of clouds is distributed quasi-uniformly in a volume of roughly $10^7$ $pc^3$, with the remaining 1 – *f* being part of the Wave. We treat *f* entirely as a nuisance parameter since it is completely degenerate with the volume of our uniform outlier model, although we have specified it such that the uniform component will contribute a "minority" of the fit (< 40%).

Assuming the distances to each of our *n* clouds have been derived independently and defining θ = $\{x_0,y_0,z_0,x_1,y_1,z_1,x_2,y_2,z_2,P,A,\phi,\gamma,\delta,\sigma,f\}$, the likelihood for a given realization of our 16-parameter 3-D model is

$$\mathcal{L}(\theta) = \prod_{i=1}^{n} \left[ (1-f)\mathcal{L}_{{\rm cloud},i}(\theta) + f\mathcal{L}_{{\rm unif},i}(\theta) \right] \quad (3)$$

where

$$\mathcal{L}_{{\rm cloud},i}(\theta) = \frac{1}{\sqrt{2\pi\sigma^2}} \exp\left[ -\frac{1}{2}\frac{d^2_{{\rm cloud},i}}{\sigma^2} \right] \;, \quad \mathcal{L}_{{\rm unif},i}(\theta) = 10^{-7} \quad (4)$$

We infer the posterior probability distribution $\mathcal{P}$(θ) of 3-D model parameters consistent with our cloud distances (excluding all "bridging" features) using Bayes' Theorem:

$$\mathcal{P}(\theta) \propto \mathcal{L}(\theta)\pi(\theta) \quad (5)$$

where π(θ) is our prior distributions over the parameters of interest. We set our prior π(θ) to be independent for each parameter, based on initial fits. The priors on each parameter are described in Extended Data Table 1, where $\mathcal{N}$(μ,σ) is a Normal distribution with mean μ and standard deviation σ and $\mathcal{U}$(a,b) is a Uniform distribution with lower bound a and upper bound b.

We generate samples from $\mathcal{P}$(θ) with the nested sampling code dynesty[32] using a combination of uniform sampling with multi-ellipsoid decompositions and 1000 live points. A summary of the derived properties of the Wave are listed in Extended Data Table 2 along with their associated 95% credible intervals (CIs). The 20 random samples from $\mathcal{P}(\theta)$ are plotted in Figure 2 to illustrate the uncertainties in our model.

Using our samples, we associate particular sightlines with the Wave by computing the mean odds ratio averaged over our posterior

$$\langle R_i \rangle = \int \frac{(1-f)\mathcal{L}_{\mathrm{cloud},i}(\theta)}{f\mathcal{L}_{\mathrm{unif},i}(\theta)} \mathcal{P}(\theta) d\theta \quad (6)$$

based on our set of samples. We subsequently classify all objects with $\langle R_i \rangle > 1$ as being part of the Radcliffe Wave, which is used as the criteria for associating sources in Figure 2. We find that this condition holds true for 43% of the sources used to determine our initial model. Our overall conclusions do not change if larger, more selective thresholds are chosen.

As further validation, we subsequently compute $\langle R_i \rangle$ for each of the 54 “bridging” sightlines targeted to follow the projected structure of the Radcliffe Wave. We find that all 54 sightlines satisfy our $\langle R_i \rangle > 1$ condition, further confirming the continuous nature of the Wave between individual clouds.

In addition to the parameters derived above, we estimate the total “length” of the feature in our dataset by computing the line integral along our model from the clouds at the endpoints, finding a length of 2.7 ± 0.2kpc (95% CI). This and other derived physical properties of the feature are highlighted in Extended Data Table 3.

---

## Methods Bibliography

---

### Data Availability

The datasets generated during and/or analyzed during the current study are publicly available on the Harvard Dataverse. The distances to the major star-forming clouds are available at https://doi.org/10.7910/DVN/07L7YZ, while the tenuous connections are available at https://doi.org/10.7910/DVN/K16GQX.

### Code Availability

The software used to determine the distances to star-forming regions is publicly available on Zenodo (https://doi.org/10.5281/zenodo.3348370 and

https://doi.org/10.5281/zenodo.3348368). The code used for the model-fitting is available from contributing author J.S. Speagle on reasonable request.

## Acknowledgements

JA is thankful to the Radcliffe Institute where this work was developed, and where he discovered the work of visual artist Anna von Mertens on Henrietta Leavitt's life's work, which inspired us to "see more." The authors acknowledge the organizers and the participants of the 2018 Paris-Saclay International Programs for Physical Sciences "The Milky Way in the age of Gaia" workshop and the Interstellar Institute for discussions at the early stage of this work. We benefited from generous discussions with Tom Dame, Mark Reid, Andreas Burkert, and Melvyn Davies. JA acknowledges the TURIS and the Data Science Research Platforms of the University of Vienna. CZ and JSS are supported by the NSF Graduate Research Fellowship Program (Grant No. 1650114) and the Harvard Data Science Initiative. DPF and CZ acknowledge support by NSF grant AST-1614941. ES acknowledges support for this work by NASA through ADAP grant NNH17AE75I and Hubble Fellowship grant HST-HF2-51367.001-A awarded by the Space Telescope Science Institute, which is operated by the Association of Universities for Research in Astronomy, Inc., for NASA, under contract NAS 5-26555. The computations in this paper utilized resources from the Odyssey cluster, which is supported by the FAS Division of Science Research Computing Group at Harvard University. The glue high-dimensional visualization software used to explore, visualize, and understand the Radcliffe wave was created by AG, TR, CZ and others, and has been supported by: US Government contract NAS5-03127 through NASA's James Webb Space Telescope Mission; NSF Award OAC-1739657 and NSF Award AST-1908419. We are grateful to Alex Johnson and others at plot.ly for their help creating the 3D interactive figure in this paper, which was output from glue to plot.ly. WorldWide Telescope (WWT), used within glue to visualize the wave, is currently supported by an NSF grant 1642446 to the American Astronomical Society. WWT was originally created by Curtis Wong and Jonathan Fay at Microsoft Research, which WWT supported development before AAS took ownership. JSS thanks Rebecca Bleich and JA thanks Anna, Julia, Matteo, and Rocco for the continuing support.

## Author Contributions

JA led the work and wrote the majority of the text. All authors contributed to the text. CZ and JSS led the data analysis and distance modeling with EFS, GMG, and DPF. CZ and JA led the kinematics analysis. JA, CZ, and AAG led the visualization efforts. JSS led the 3-D modeling. JA, CZ, and AAG led efforts to interpret the results. TR, AAG, JSS, and CZ contributed to software used in this work.

## Author information

Correspondence and requests for materials should be addressed to JA (email: joao.alves@univie.ac.at).

## Competing Interests

The authors declare that they have no competing financial interests.

# Extended Data Tables.

Extended Data Table 1: Priors on Radcliffe Wave Parameters

| Parameter | Prior | Parameter | Prior |
|---|---|---|---|
| $x_0$ | $\mathcal{N}\,(-900\,\mathrm{pc}, 100\,\mathrm{pc})$ | $P$ | $\mathcal{N}\,(3500\,\mathrm{pc}, 300\,\mathrm{pc})$ |
| $y_0$ | $\mathcal{N}\,(-900\,\mathrm{pc}, 100\,\mathrm{pc})$ | $A$ | $\mathcal{N}\,(170\,\mathrm{pc}, 20\,\mathrm{pc})$ |
| $z_0$ | $\mathcal{N}\,(0\,\mathrm{pc}, 50\,\mathrm{pc})$ | $\phi$ | $\mathcal{N}\,(2.9\,\mathrm{rad}, 0.5\,\mathrm{rad})$ |
| $x_1$ | $\mathcal{N}\,(-300\,\mathrm{pc}, 100\,\mathrm{pc})$ | $\ln\gamma$ | $\mathcal{N}\,(-0.5, 0.5)$ |
| $y_1$ | $\mathcal{N}\,(0\,\mathrm{pc}, 100\,\mathrm{pc})$ | $\ln\delta$ | $\mathcal{N}\,(-0.5, 0.5)$ |
| $z_1$ | $\mathcal{N}\,(0\,\mathrm{pc}, 50\,\mathrm{pc})$ | $\ln\sigma/\mathrm{pc}$ | $\mathcal{U}\,(3.5, 5)$ |
| $x_2$ | $\mathcal{N}\,(300\,\mathrm{pc}, 100\,\mathrm{pc})$ | $f$ | $\mathcal{U}\,(0.15, 0.4)$ |
| $y_2$ | $\mathcal{N}\,(1400\,\mathrm{pc}, 100\,\mathrm{pc})$ | | |
| $z_2$ | $\mathcal{N}\,(0\,\mathrm{pc}, 50\,\mathrm{pc})$ | | |

Extended Data Table 2: Constraints on Radcliffe Wave Parameters

| Parameter | Median with 95% CI |
|---|---|
| $x_0$ | $-910^{+130}_{-150}\,\mathrm{pc}$ |
| $y_0$ | $-860^{+140}_{-130}\,\mathrm{pc}$ |
| $z_0$ | $-30^{+80}_{-80}\,\mathrm{pc}$ |
| $x_1$ | $-270^{+80}_{-80}\,\mathrm{pc}$ |
| $y_1$ | $-20^{+160}_{-150}\,\mathrm{pc}$ |
| $z_1$ | $-10^{+30}_{-30}\,\mathrm{pc}$ |
| $x_2$ | $290^{+70}_{-70}\,\mathrm{pc}$ |
| $y_2$ | $1400^{+170}_{-170}\,\mathrm{pc}$ |
| $z_2$ | $30^{+50}_{-50}\,\mathrm{pc}$ |

| Parameter | Median with 95% CI |
|---|---|
| $P$ | $3560^{+500}_{-470}\,\mathrm{pc}$ |
| $A$ | $160^{+30}_{-30}\,\mathrm{pc}$ |
| $\phi$ | $2.89^{+0.51}_{-0.58}\,\mathrm{rad}$ |
| $\gamma$ | $0.50^{+0.27}_{-0.17}$ |
| $\delta$ | $6.68^{+5.48}_{-3.20}$ |
| $\sigma$ | $62^{+16}_{-13}\,\mathrm{pc}$ |
| $f$ | $0.22^{+0.07}_{-0.06}$ |

Extended Data Table 3: Physical Properties of the Radcliffe Wave

| Name | Median with 95% CI |
|---|---|
| Length | $2.7 \pm 0.2$ kpc |
| Scatter | $60 \pm 15$ pc |
| Amplitude | $160 \pm 30$ pc |
| Mass | $\geq 3 \times 10^6$ $M_\odot$ |

## Extended Data Figures.

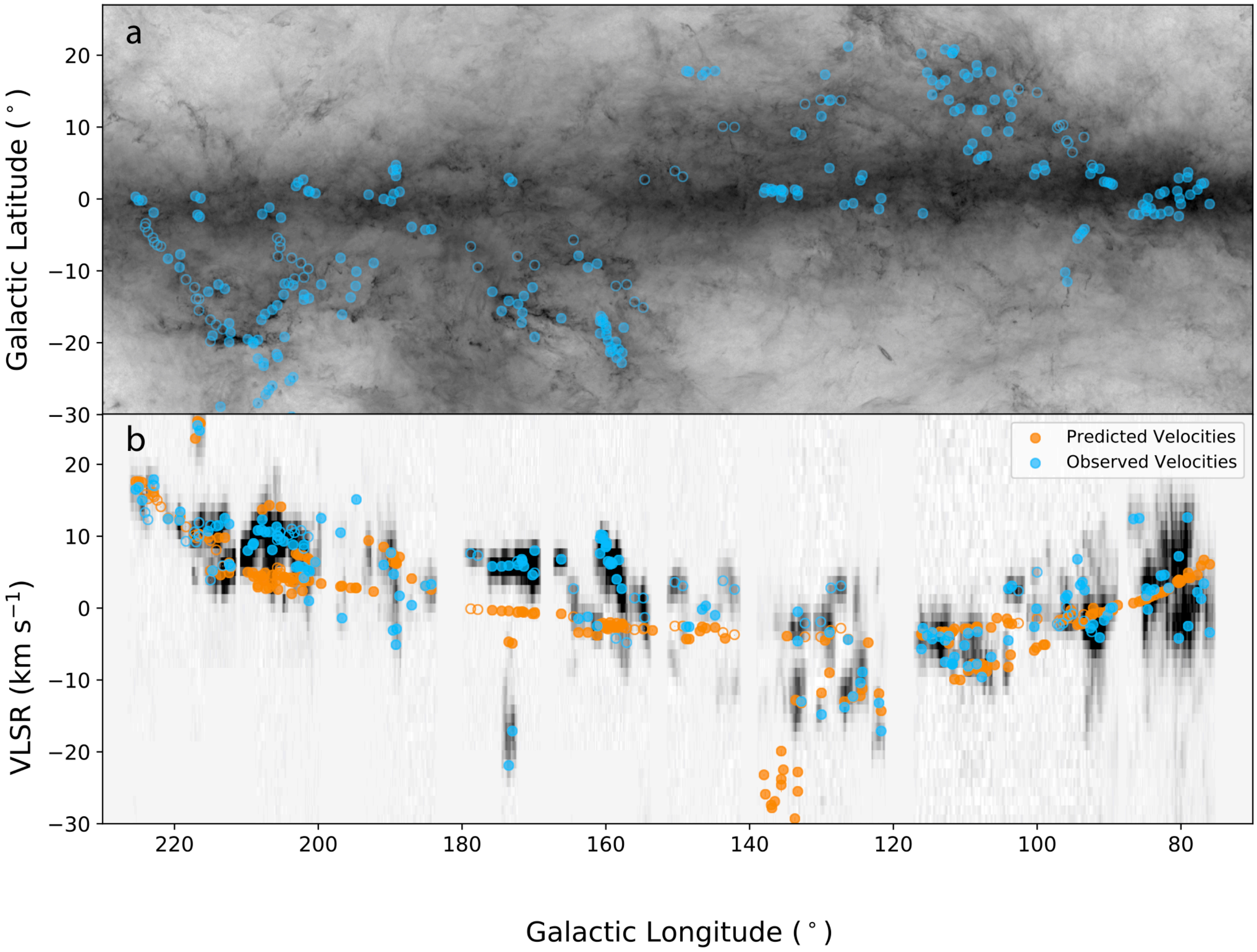

Extended Data Figure 1. Position-velocity diagram. The blue points in panel a. are as in Figure 1 while the orange points in panel b. represent the predicted positions of the blue points as if they were following a “universal” Galactic rotation curve[22]. 1σ errors—derived from the Gaussian fitting for the observed velocities and the distance uncertainties for the predicted velocities—are shown via the line segments, but are generally smaller than the symbols. The quasi-linear arrangement in velocity of the Radcliffe Wave complexes (labeled) suggests that the new structure is not a random alignment of molecular cloud complexes, but a kinematically coherent structure. The tentative decoupling between observed and predicted velocities also indicate that the Wave is a kinematically coherent structure.